\documentclass[mypaper,7pt,twoside]{CoAst}
\usepackage{epsf,graphicx,fancyhdr,lscape,psfig}
\renewcommand{\chaptermark}[1]{\markboth{#1}{}}

\newcommand{\kopf}{\small\itshape Comm. in Asteroseismology\\ Vol. 148, 2006}
\newcommand{\Authors}[1]{\begin{center}\normalsize\bf\sf #1 \end{center}}

\renewcommand{\author}[1]{\begin{center}\normalsize\bf\sf #1 \end{center}}
\newcommand{\Address}[1]{\begin{center}\small\sf #1 \end{center}}

\newcommand{\References}[1]{\begin{flushleft}{\large References\\}\vspace*{2mm}\small #1 \end{flushleft}}

\newcommand{\chapterDSSN}[2]{\chapter[\sf\normalsize #1\\ \footnotesize \hspace*{5mm}by #2 \sf\normalsize][]{#1\\}\rhead[\fancyplain{}{\sf\footnotesize \center{#1}}]{\fancyplain{}{\sffamily\thepage}}\lhead[\fancyplain{\kopf}{\sffamily\thepage}]{\fancyplain{\kopf}{\sf\footnotesize \center{#2}}}}

\renewcommand{\chaptername}{}

\begin{document}
\sf

\chapterDSSN{Time resolved spectroscopy of the multiperiodic pulsating subdwarf B star PG1605+072}{A. Tillich, U. Heber and S. J. O'Toole}

\Authors{A. Tillich$^1$, U. Heber$^1$ and S. J. O'Toole$^2$} 
\Address{$^1$Dr.Remeis-Sternwarte Bamberg, Universit\"at Erlangen-N\"urnberg, Sternwartstr.7, D-96049 Bamberg, Germany\newline$^2$Anglo-Australian Observatory, P.O. Box 296 Epping, NSW 1710, Australia}

\noindent
We present results for the 2m spectroscopic part of the
MultiSite Spectroscopic Telescope campaign, which took place in May/June
2002. In order to perform an asteroseismological analysis on the multiperiodic
pulsating subdwarf B star PG 1605+072 we used over 150 hours of time
resolved spectroscopy to search for and analyse line profile variations by using
phase binning.
This pilot analysis using the \textit{BRUCE} and \textit{KYLIE} programs and
assuming strong rotation and low inclination favours models with $l=1$ or $l=2$ with $m\leq0$.





\section{The MSST data and phasebinning}
Four observatories (Steward Observatory, ESO, Siding Spring Observatory, NOT) produced 10892 time resolved spectra. O'Toole et al.(2005) detected the 20 strongest modes in radial velocity. Here we treat the data sets of each telescope separately. After reducing the spectra using IRAF we coadded them according to their phase for the four dominant modes. Then we fitted LTE-model spectra using the \textit{FITPROF} program (Napiwotzki, 1999) in order to determine simultaneously the three atmospheric parameters for every bin. The results derived from the Steward Observatory data are shown in Fig. 1a,b.
 \begin{figure}[h]
\begin{center}
 \includegraphics [scale=.22] {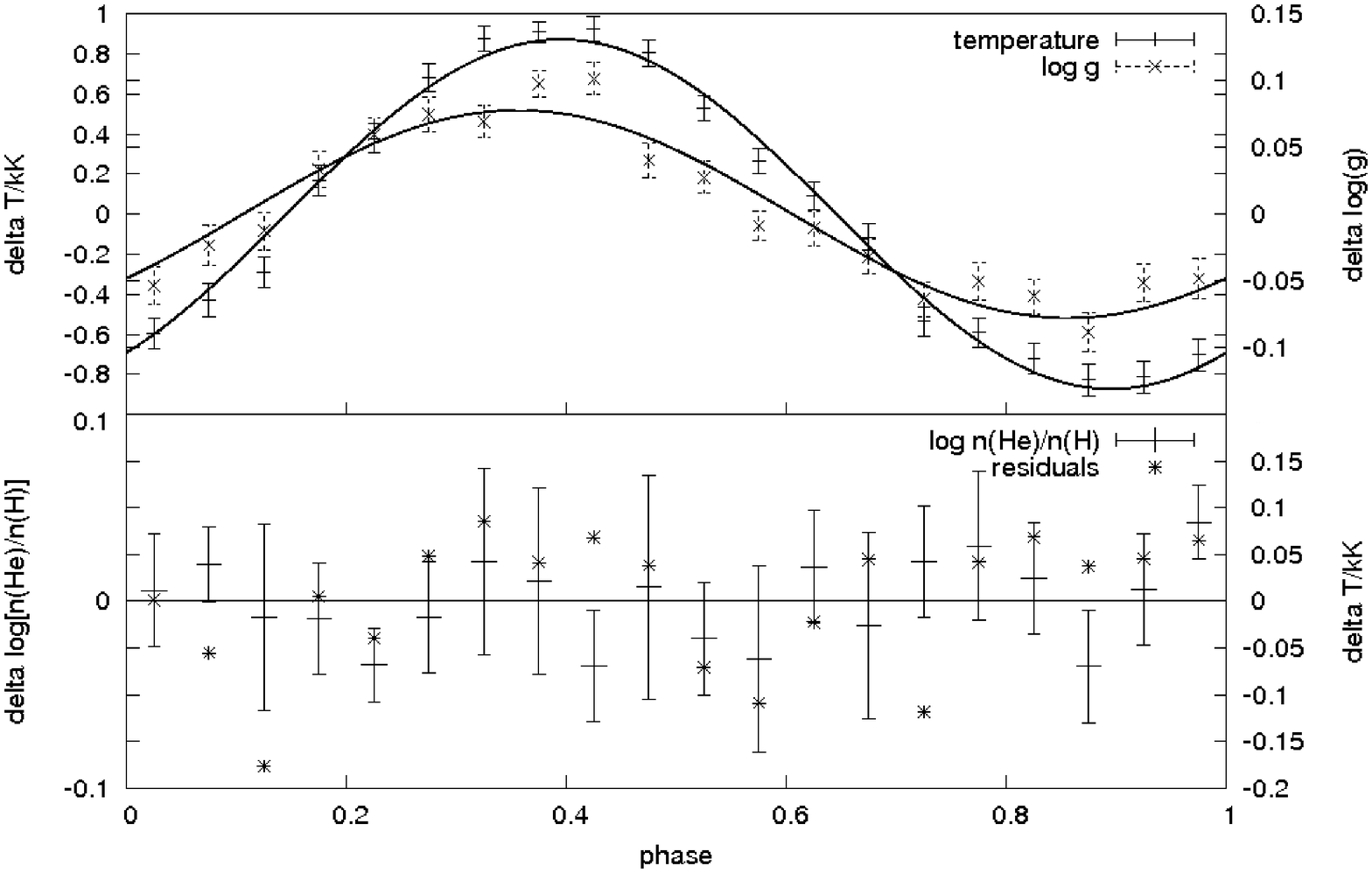}
 \includegraphics [scale=.22] {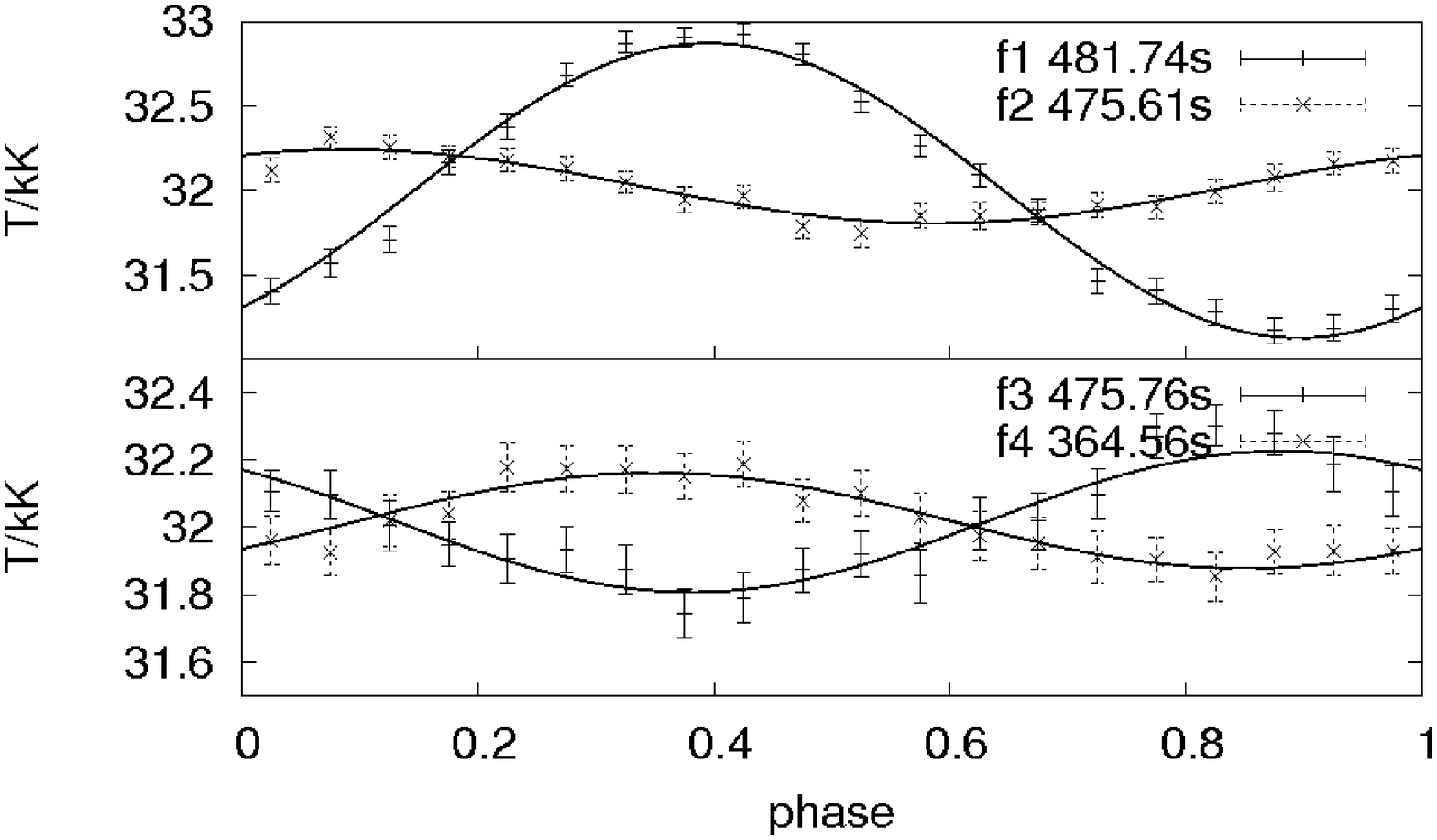}
\end{center}
\caption{\textit{a (left panel)}: variations of the atmospheric values with errors and sine fit for the strongest mode f1 (481.74s).\newline\textit{b (right panel)}: temperature variations for the four strongest modes with errors and sine fit.}
 \end{figure}

\section{Modelling of line profile variations and mode identification}
In order to identify the modes, we used the \textit{BRUCE} and \textit{KYLIE} routines (R. Townsend, 1997), to model various pulsation modes by perturbing our static models. We then determined the atmospheric parameters of the perturbed models using \textit{FITPROF}. The required parameters ($v_{rot}^{eq}=130\mathrm{kms^{-1}}$,$i=17^\mathrm{\circ}$) were taken from previous analyses (Heber et al., 1999; Kawaler, 1999). The results are shown in Fig. 2a,b.
 \begin{figure}[h]
\begin{center}
 \includegraphics [scale=.22] {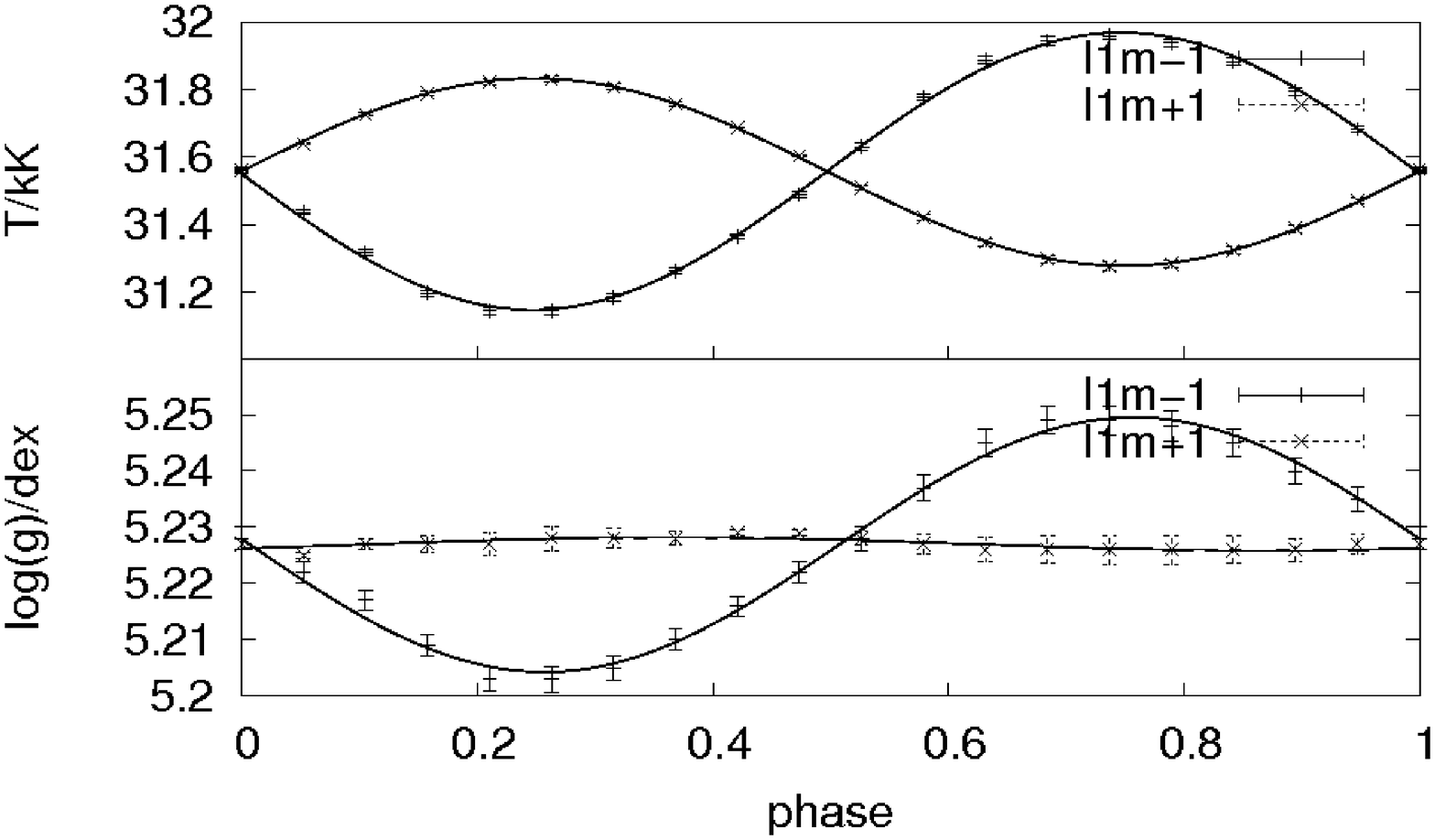}
 \includegraphics [scale=.22] {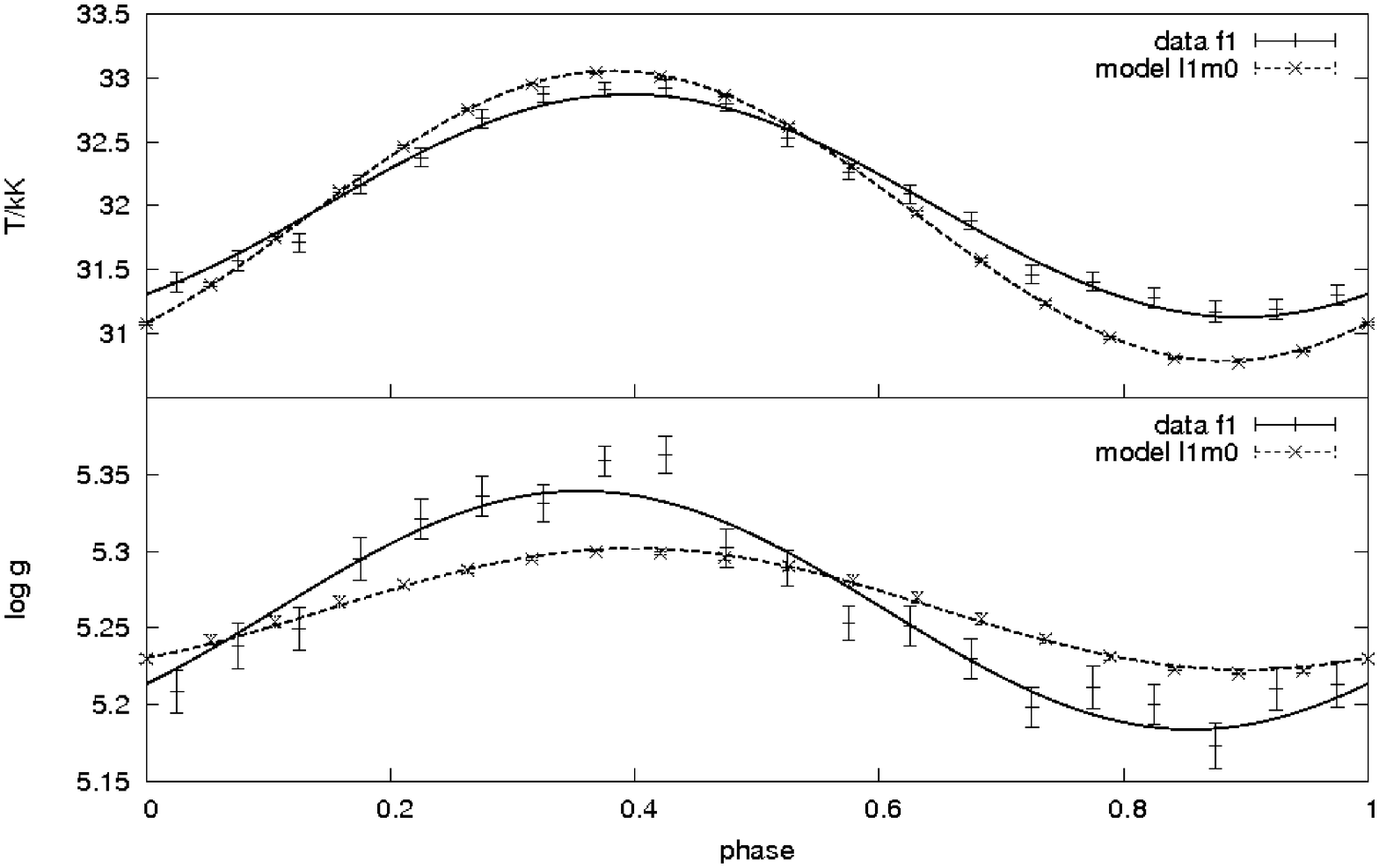}
\end{center}
\caption{ variation of $T_{\mathrm{eff}}$ and $\log{g}$ with errors and sine fit\newline\textit{a (left panel)}: models for $l=1,m=\pm1$.\newline\textit{b (right panel)}: observations of the dominant period and model $l=1,m=0$.}
 \end{figure}
For the the dominant period (481.74s) the mode with l=1,m=0 is the best fit. But also for the other three modes, this pilot analysis assuming strong rotation and low inclination favours models with $l=1$ or $l=2$ with $m\leq0$. Nevertheless the parameter range has to be further exploited to derive a consistent model.\\


\References{
Heber, U., Reid, I.N. \& Werner K. et al., 1999, A\&A, 348, L25.\\
Kawaler, S., 1999, in 11th. european Workshop on White Dwarfs, ASPC169, 158.\\
Napiwotzki, R., 1999, A\&A, 350, 101.\\
O'Toole, S. J., Heber, U., Jeffery, C.S. et al., 2005, A\&A, 440, 667.\\
Townsend, R., 1997, PhD Thesis, University College London.\\
}

\end{document}